\documentclass[9pt,twocolumn,twoside]{osajnl}
\usepackage{bm}
\usepackage{color}
\usepackage{mathrsfs}
\journal{ol} 

\setboolean{shortarticle}{false}

\ifthenelse{\boolean{shortarticle}}{\colorlet{color2}{color2b}}{\colorlet{color2}{color2}} 

\title{Enhancement of spatial resolution of ghost imaging via localizing and thresholding}

\author[1*]{Yunlong Wang}
\author[1*]{Yingnan Zhou}
\author[1]{Shaoxiong Wang}
\author[1]{Feiran Wang}
\author[1**]{Ruifeng Liu}
\author[1]{Pei Zhang}
\author[1]{Hong Gao}
\author[1**]{Fuli Li}

\affil[1]{Shaanxi Province Key Laboratory for Quantum Information and Quantum Optoelectronic Devices, and Department of Applied Physics, School of Science,
Xi'an Jiaotong University, Xi'an 710049, People's Republic of China}
\affil[*]{These authors contributed equally to this work.}
\affil[**]{Co-Corresponding author: ruifeng.liu@mail.xjtu.edu.cn, flli@mail.xjtu.edu.cn}

\dates{Compiled \today}

\ociscodes{(100.2980) Image enhancement; (100.6640) Superresolution; (270.0270) Quantum optics.}

\doi{\url{http://dx.doi.org/10.1364/ao.XX.XXXXXX}}

\begin{abstract}

In ghost imaging scheme, an illuminated light is split into test and reference beams which pass through two different optical systems respectively and an image is constructed by the second-order correlation between the two light beams. Since both the two light beams are all diffracted when passing through the optical systems,
spatial resolution of ghost imaging is in general lower than that of a corresponding conventional imaging system.
When Gaussian-shaped light spots are used to illuminate an object, randomly scanning across the object plane, in ghost imaging scheme, we show that by localizing central positions of the spots of the reference light beam, the resolution can be enhanced by a factor of $\sqrt{2}$ same as that of the corresponding conventional imaging system. We also find that the resolution can be further enhanced by setting an appropriate threshold to the bucket measurement of ghost imaging.

\end{abstract}

\setboolean{displaycopyright}{true}

\begin{document}

\maketitle
\thispagestyle{fancy}

\ifthenelse{\boolean{shortarticle}}{\ifthenelse{\boolean{singlecolumn}}{\abscontentformatted}{\abscontent}}{}

\section{Introduction}

Different from the conventional imaging which is based on the light's first-order correlation, the correlated imaging including auto-correlation imaging scheme \cite{gaojie2,Oh2013,long,Zhang2015} and ghost imaging scheme \cite{gm2,clslit,hot,incoherent,liu2016enhanced} employs the light's second-order correlation.
It results in some unique features and has attracted a lot of attention in the last two decades. In fact, correlated imaging has been applied in various fields including optical lithography \cite{shike1,liu2014super}, remote imaging \cite{remote}, microscopy imaging \cite{shengwu2}, X-ray imaging \cite{Pelliccia2016}, and Terahertz imaging \cite{Stantchev2016}.


In those applications, one of the important issues is the image's spatial resolution.
In conventional imaging, the resolution is given by Rayleigh criterion \cite{LordRay} which comes from the finite-size of the imaging system's entrance pupil.
In the auto-correlation imaging system, the resolution is determined by a product of point spread functions (PSFs) of the two light beams \cite{gaojie2,Oh2013,long,Zhang2015}.
In ghost imaging scheme, the resolution is in general quantified by the full width at half-maximum (FWHM) of the peak in a Hanbury Brown and Twiss (HBT) intensity correlation measurement, or by use of the transverse coherence length of the speckles of illumination light, 
namely the averaged size of the speckles \cite{gaojie2,Ferri2005,Ferri2008,Chen17}.
In the last decade, a great deal of resolution enhancement proposals have been suggested, such as compressive sensing technique \cite{post1,xishu1,xushu2} and non-Rayleigh speckle fields \cite{Kuplicki2016,Zhang2016}, low-pass spatial filter scheme \cite{Chen17} and high-pass spatial-frequency filter scheme \cite{Sprigg2016}.
In a very recent publication, Yang et al. \cite{Yang2017} reported a new correlated imaging scheme using the orbital angular momentum correlations of light. This scheme may be used to image an object with a very high azimuthal resolution.
Although those investigations to the resolution of ghost imaging have been done, to our knowledge, how the optical systems in the test and the reference path of ghost imaging individually affect the imaging resolution is not clear. Is the resolution of ghost imaging also given  by a product of the two light beam's PSFs as in the auto-correlation imaging scheme?

In this letter, a Gaussian-shaped spot illuminated model is employed to theoretically analyze the resolution issue of ghost imaging. It is noticed that the resolution of ghost imaging is determined by the convolution of the PSFs of the test and reference light beams. Then, we propose localization and threshold methods to
narrow the PSFs and enhance the resolution.
In the photoactivated localization microscopy (PALM) scheme \cite{betzig2006imaging}, a sub-diffraction image is constructed by localizing the positions of the fluorophores \cite{cheezum2001quantitative,betzig2006imaging}. Inspired by PALM, we show that by localizing the central positions of all the illuminated spots of the reference light beam, the resolution of ghost imaging can be enhanced by a factor of $\sqrt{2}$.
We also find that the resolution can be further enhanced by setting an appropriate threshold  to the bucket measurement \cite{sp1,sp2,post2}.

\section{Spatial resolution of ghost imaging}

\begin{figure}[htbp]
\centering
\includegraphics[width=6 cm]{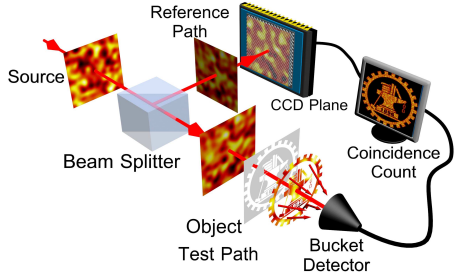}
\caption{Schematic of ghost imaging. An illumination light is split at a 50:50 beam splitter into two paths. In the test path, the light illuminates the object and then is collected by a bucket detector with no spatial resolution. In the reference path, the light intensity distribution is recorded by a charge-coupled device (CCD) camera.}
\label{fig1}
\end{figure}

A typical ghost imaging schematic is shown in Fig. \ref{fig1}. Suppose that
the light beam in the test path propagates to the object through an optical system with a PSF $h_t(\bm{\rho_t}-\bm{\rho})$.
Here $\bm{\rho}$ and $\bm{\rho_t}$ are the transverse coordinates on the source and the object plane, respectively.
If $T(\bm{\rho_t})$ represents the transmission function of the object, the light field behind the object is given by

\begin{equation}
E_t^{(+)}(\bm{\rho_t})=\int d\bm{\rho}T(\bm{\rho_t})h_t(\bm{\rho_t}-\bm{\rho})E_s^{(+)}(\bm{\rho}),
\label{eq101}
\end{equation}
where $E_s^{(+)}(\bm{\rho})$ denotes the positive frequency part of the source field at $\bm{\rho}$.
The optical-electric current operator at the bucket detector is
\begin{align}
I_t&=\int d\bm{\rho_t} E_t^{(+)}(\bm{\rho_t}) E_t^{(-)}(\bm{\rho_t}).
\label{eq102}
\end{align}

Suppose that the PSF of the reference path is $h_r(\bm{\rho_r}-\bm{\rho})$.
The optical field on the charge-coupled device (CCD) plane is given by
\begin{equation}
E_r^{(+)}(\bm{\rho_r})=\int d\bm{\rho}h_r(\bm{\rho_r}-\bm{\rho})E_s^{(+)}(\bm{\rho}).
\label{eq103}
\end{equation}
Thus, the photon count operator at the CCD plane is
\begin{align}
I_r(\bm{\rho_r})&=E_r^{(+)}(\bm{\rho_r}) E_r^{(-)}(\bm{\rho_r}).
\label{eq1033}
\end{align}

If the light source is in a Gaussian-type quantum state, the fluctuation of the coincidence count can be written as \cite{liu2016enhanced}
\begin{align}
\Delta C(\bm{\rho_r})&=\left\langle I_t-\left\langle I_t\right\rangle\right\rangle\left\langle I_r-\left\langle I_r\right\rangle\right\rangle
\nonumber \\
&=\int d\bm{\rho_t}|T(\bm{\rho_t})|^2 \int d\bm{\rho_1} d\bm{\rho_2}n_s(\bm{\rho_1})n_s(\bm{\rho_2})
\nonumber \\
&\times h_t(\bm{\rho_t}-\bm{\rho_1}) h_t^*(\bm{\rho_t}-\bm{\rho_2})  h_r(\bm{\rho_r}-\bm{\rho_2}) h_r^*(\bm{\rho_r}-\bm{\rho_1}),
\label{eq1010}
\end{align}
where $n_s(\bm{\rho})$ represents the mean photon number of the source.

In order to explicitly analyze spatial resolution of ghost imaging, we consider the case where the illuminated light is a point-like source which randomly and uniformly distributed on the source plane.
If the light spot is located at $\bm{\rho_0}$, we may have $n_s(\bm{\rho})=n_{sc}\delta(\bm{\rho}-\bm{\rho_0})$, where $n_{sc}$ is a constant. In this case, we have
\begin{align}
\Delta C(\bm{\rho_r})&=n_{sc}^2\int d\bm{\rho_0}d\bm{\rho_t}|T(\bm{\rho_t})|^2| h_t(\bm{\rho_t}-\bm{\rho_0})|^2  |h_r(\bm{\rho_r}-\bm{\rho_0})|^2 \nonumber \\
&=n_{sc}^2 |T(\bm{\rho_r})|^2\otimes |h_t(\bm{\rho_r})|^2 \otimes |h_r(\bm{\rho_r})|^2.
\label{eq1011}
\end{align}

As is well known, the image in the conventional imaging system can be expressed in the form \cite{shu2}
\begin{equation}
I_{ci}(\bm{\rho_i})=|T(\bm{\rho_i})|^2 \otimes |h(\bm{\rho_i})|^2,
\label{eq:refname 10}
\end{equation}
where $\bm{\rho_i}$ is the transverse coordinate on the image plane and $h(\bm{\rho_i})$ stands for the PSF of the imaging system.
Obviously, the image resolution is confined by this PSF. In general, it is a Sombrero function that produces the Airy disk on the image plane and the first zero of Sombrero function leads to the resolution limit.
Note that Eq. (\ref{eq1011}) has a similar form as Eq. (\ref{eq:refname 10}).
It means that the convolution function $\mathcal{H}(\bm{\rho_r}) = |h_t(\bm{\rho_r})|^2 \otimes |h_r(\bm{\rho_r})|^2$ in Eq. (\ref{eq1011})
behaves the same effect as the PSF $|h(\bm{\rho_i})|^2$ in Eq. (\ref{eq:refname 10}).

In order to work out an explicit expression of the resolution of ghost imaging, we assume that all the PSFs in Eqs. (\ref{eq1011}) and (\ref{eq:refname 10})
have the same Gaussian form $\exp\left[-\frac{{\bm{\rho}}^2}{2\sigma ^2}\right]$.
Inserting the PSF into Eq. (\ref{eq1011}) and (\ref{eq:refname 10}), we obtain the formula for ghost imaging
\begin{equation}
\Delta C(\bm{\rho_r})=n_{sc}^2 \int d\bm{\rho_t}|T(\bm{\rho_t})|^2 e^{-\dfrac{(\bm{\rho_t}-\bm{\rho_r})^2}{2\sigma^2}},
\label{eq16}
\end{equation}
and the formula for conventional imaging
\begin{equation}
I_{ci}(\bm{\rho_i})= \int d\bm{\rho_t}|T(\bm{\rho_t})|^2 e^{-\dfrac{(\bm{\rho_t}-\bm{\rho_i})^2}{\sigma^2}}.
\label{eq17}
\end{equation}
In Eq. (\ref{eq17}), the FWHM of the function $|h(\bm{\rho_i})|^2=\exp\left[-\dfrac{(\bm{\rho_t}-\bm{\rho_i})^2}{\sigma^2}\right]$ is $2\sqrt{\ln{2}}\sigma$.
As for ghost imaging, due to the convolution of the two Gaussian functions in Eq. (\ref{eq1011}), the FWHM of the total PSF,  $\mathcal{H}(\bm{\rho_r})=\exp\left[-\dfrac{(\bm{\rho_t}-\bm{\rho_r})^2}{2\sigma^2}\right]$, is $2\sqrt{2\ln{2}}\sigma$ which is $\sqrt{2}$ wider than that of the PSF in Eq. (\ref{eq17}).
It means that if the optical systems used for both conventional imaging and ghost imaging consist of the same entrance pupil, the resolution of ghost imaging is lower than that of the conventional imaging by a factor of $\sqrt{2}$.

Now, it becomes clear that the resolution of ghost imaging can be improved if the FWHMs of the two PSF functions in $\mathcal{H}(\bm{\rho_r})$ is reduced.
For this end, as similar as in the localization process \citep{betzig2006imaging}, we precisely label central positions of light spots of each illumination pattern in the reference light beam. In mathematics, it means that the PSF $h_r(\bm{\rho_r}-\bm{\rho_0})$ of the reference beam in Eq. (\ref{eq1011}) is replaced by a $\delta$-function $\delta(\bm{\rho_r}-\bm{\rho_0})$ \cite{cheezum2001quantitative,betzig2006imaging}.
In this way, Eq. (\ref{eq1011}) can be simplified as
\begin{align}
\Delta C(\bm{\rho_r})&=n_{sc}^2 |T(\bm{\rho_r})|^2\otimes |h_t(\bm{\rho_r})|^2\label{eq:refname 11}
\\ \nonumber
&=n_{sc}^2 \int d\bm{\rho_t}|T(\bm{\rho_t})|^2 e^{-\dfrac{(\bm{\rho_t}-\bm{\rho_r})^2}{\sigma^2}}.
\end{align}
As the imaging formulae, obviously, Eq. (\ref{eq17}) and Eq. (\ref{eq:refname 11}) give the same spatial resolution.
It means that the resolution of ghost imaging with localizing the spot positions can be the same as that of the corresponding conventional imaging.

\begin{figure}[htbp]
\centering
\includegraphics[width=7 cm]{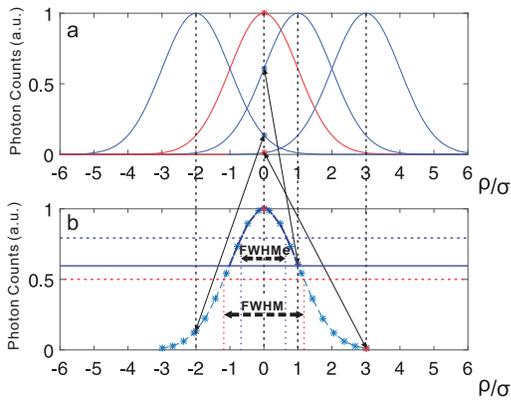}
\caption{(a) A pinhole-like object located at $\bm{\rho}=0$ is illuminated by single Gaussian-shaped spots with different central locations.
(b) Image of the pinhole. The blue solid line indicates a threshold to the bucket measurement.}
\label{fig:explain}
\end{figure}

Next, let us discuss how to further improve the resolution of ghost imaging. From Eq. (10), we see that after localizing the spot positions, the resolution is now determined by the PSF $h_t(\bm{\rho_r})$ of the test path. In order to understand how the PSF affects the resolution, we study the case where the object is a pinhole at $\bm{\rho}=0$ as shown in Fig. \ref{fig:explain}(a). When one Gaussian-shaped spot shines on the object, the corresponding bucket measurement value is proportional to the light intensity at the location of the pinhole and varies with change of the central position of the light spot. To implement the second-order correlation process in ghost imaging, we assign  the bucket measurement value to the central position of the spot in the reference path. After the ensemble average of all the illuminated spots, as shown in Fig. \ref{fig:explain}(b), we obtain the Gaussian-shaped curve which is the image of the pinhole. This result can easily be obtained by substituting a $\delta$-function as a transmission function of the object in Eq. (10). In Fig. \ref{fig:explain}(b), the FWHM of the image curve is labeled by the red dashed double-arrowed lines. Here, we choose the post-selection method \cite{post2} to reduce this FWHM. With this method, only the bucket measurement values larger than a threshold are reserved and others are abandoned. If the maximal value of the bucket measurement is $N_{max}$ and the threshold is set to be $N_{th}$, as shown in Fig. \ref{fig:explain}(b), the effective FWHM labeled by the blue dashed double-arrowed lines is given by $2\sqrt{\ln\frac{2N_{max}}{N_{max}+N_{th}}}\sigma\leq 2\sqrt{\ln{2}}\sigma$. Thus, the resolution of ghost imaging can be enhanced when an appropriate threshold is set to the bucket measurement.

\section{experiment setup and results}

\begin{figure}[htbp]
\centering
\includegraphics[width=8 cm]{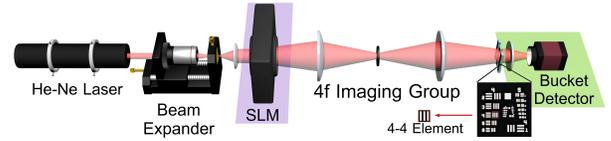}
\caption{Experimental setup of computational ghost imaging. The beam expander is used to expand the transverse section of the light beam from the laser source. The Spatial Light Modulator (SLM) is controlled by a computer to generate illuminated patterns of light spots with random central positions. After demagnified by the 4f-system, the light spots illuminate the object (Group 4, Element 4 portion of USAF resolution chart), and the transmitted light is collected by a bucket detector.}
\label{fig:1}
\end{figure}

The experimental setup is shown in Fig. \ref{fig:1}.
A light beam from a He-Ne laser with wavelength $\lambda=632.8$ nm passes through the beam expander and arrives at the Spatial Light Modulator (SLM, RSLM1024U, Shanghai Realic Information Technology Corporation,
Ltd). The SLM is codified by a computer to generate the transverse patterns with random distribution of light spots.
After the SLM, the light passes through a 4f-system and then it is demagnified by a factor of 6. The Group 4, Element 4 portion of the resolution chart (Thorlabs negative USAF1951), which consists of triple alternating stripes with the separation of 62.50 $\mu$m, is taken as an object. The transmitted light of the object is collected by a bucket detector. In this scheme, the computer-programming patterns are recorded prior to the coincidence measurement and the computational ghost imaging is implemented \cite{jisuangui,Sun2012}.

\begin{figure}[htbp]
\centering
\includegraphics[width=6 cm]{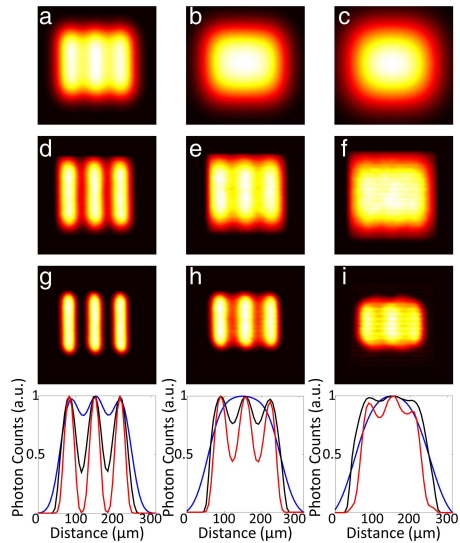}
\caption{The images obtained from the ghost imaging experimental setup described in Fig. \ref{fig1}.
From left to right column images,
the FWHM of the illuminated light spots is 44.34 $\mu m$, 64.50 $\mu m$ and 74.38 $\mu m$, respectively.
(a)-(c) the results from the conventional ghost imaging; (d)-(f) from the ghost imaging with the localization method; (g)-(i) with the localization and threshold methods by setting $N_{th}=0.6N_{max}$.
In the bottom of each column, the corresponding cross-section images are given.
The blue, black and red lines stand for the results from the conventional ghost images, the ghost imaging with the localization method and the localized ghost imaging with the $0.6N_{max}$ threshold selection, respectively.}
\label{fig:2}
\end{figure}

In experiment, the FWHM of the Gaussian-shaped spot shinning the object plane varies from 44.34 $\mu m$ to 64.50 $\mu m$, and to 74.38 $\mu m$.
The total number of the frames used to reconstruct the ghost image is 500,000.
As indicated in Eq. (\ref{eq1011}), the maximal FWHM of the spots which can distinguish the strips is 44.34 $\mu m$, because $44.34\times\sqrt{2}=62.70$ $ \mu m$ is close to the separation of the stripes.
When the FWHM takes the critical value, the result is shown in Fig. \ref{fig:2}(a) and the stripes are not very clear.
In Fig. \ref{fig:2}(b) and (c), we observe that the images become further blurred and indistinguishable with increasing the FWHM values.
The ghost images with the localization method are shown in Figs. \ref{fig:2}(d)-(f).
In constructing the images, we use a Gaussian-shaped light spot with 12 $\mu m$ FWHM as the light spots of the reference beams.
The resolution enhancement is obvious and the images become clearer not only with the FWHM 44.34 $\mu m$ but also with 64.50 $\mu m$ which is larger than the separation of the strips.

To further raise the resolution, we set a threshold $0.6N_{max}$ to the bucket measurement. It means that only the values of the bucket measurement larger than the threshold are used to construct the images.
The results are shown in Figs. \ref{fig:2}(g)-(i). By comparing Figs. \ref{fig:2}(g)-(i) with Figs. \ref{fig:2}(d)-(f), we observe that the resolution is obviously enhanced and the stirps becomes distinguishable even with the light spots of the FWHM 74.38 $\mu m$ as shown in Fig. \ref{fig:2}(i).

Finally, we would like to emphasize two points.
At first, the threshold method may also be suitable to a gray-scale object although the binary object is used in the present demonstration.
For a gray-scaled object, light spots shinning on the low gray-scale regions would be abandoned if the threshold is set too high,
and the contrast resolution would be decreased and the image get distorted.
For a gray-scaled object, one should appropriately divide the object into several zones according to its gray distribution and set different threshold values for different zones. Then, the ghost imaging process is performed zone-by-zone and the whole image of the object can be constructed by combining the images of the different zones. It means that for a gray-scale object one should dynamically set a threshold value to the bucket measurement.
Secondly, one may realize that the resolution can be enhanced if higher-order powers of the PSFs are used in Eq. (6) as in the auto-correlation imaging scheme \cite{gaojie2,Oh2013,long,Zhang2015}. In our scheme, the localizing process means that the PSF of the reference path in Eq. (6) is replaced by a $\delta$-function and the PSF is infinitely narrowed.  Thus, the localization method is equivalent to taking the infinitely high-order power of the PSF of the reference path in Eq. (6).

\section{conclusion}

In this letter, we theoretically analyzed how the PSFs of the test and the reference paths respectively affect spatial resolution of ghost imaging.
It is noticed that spatial resolution of ghost imaging is determined by the convolution of the two PSFs.
We showed that the influence of the PSF of the reference path on the resolution can totally be canceled by localizing central positions of the spots of the illumination patterns in the reference path and the image resolution can be enhanced by a factor of $\sqrt{2}$. In addition,
the influence from the PSF of the test path can be depressed by setting an appropriate threshold to the bucket measurement and
the resolution can be further enhanced.
We experimentally demonstrated the resolution enhancement, and the results are well in agreement with the theoretical prediction.

\section{Funding Information}
This work was supported by Ministry of Science and Technology of the Peoples Republic of China (MOST) (2016YFA0301404), the National Natural Science Foundation of China (NSFC) (11534008, 11374239, 11605126), the Project is also supported by Natural Science Basic Research Plan in Shaanxi Province of China (Program No. 2017JQ1025), Doctoral Fund of Ministry of Education of China (Grant 2016M592772), and the Fundamental Research Funds for the Central Universities.

\bigskip
\noindent

\bibliography{sample}

\ifthenelse{\equal{\journalref}{ol}}{%
\clearpage
\bibliographyfullrefs{sample}
}{}


\ifthenelse{\equal{\journalref}{aop}}{%
\section*{Author Biographies}
\begingroup
\setlength\intextsep{0pt}
\begin{minipage}[t][6.3cm][t]{1.0\textwidth} 
 \begin{wrapfigure}{L}{0.25\textwidth}
 \includegraphics[width=0.25\textwidth]{john_smith.eps}
 \end{wrapfigure}
 \noindent
 {\bfseries John Smith} received his BSc (Mathematics) in 2000 from The University of Maryland. His research interests include lasers and optics.
\end{minipage}
\begin{minipage}{1.0\textwidth}
 \begin{wrapfigure}{L}{0.25\textwidth}
 \includegraphics[width=0.25\textwidth]{alice_smith.eps}
 \end{wrapfigure}
 \noindent
 {\bfseries Alice Smith} also received her BSc (Mathematics) in 2000 from The University of Maryland. Her research interests also include lasers and optics.
\end{minipage}
\endgroup
}{}

\end{document}